\documentclass[prl,superscriptaddress,twocolumn,footinbib]{revtex4-2}
\pdfoutput=1
\usepackage[colorlinks=true,urlcolor=blue]{hyperref}
\usepackage{amsfonts}
\usepackage{graphicx}
\usepackage{placeins}
\usepackage{amsmath}
\usepackage{xcolor}
\usepackage{color}
\usepackage{bbold}
\usepackage{float}
\usepackage{bm}

\definecolor{red}{rgb}{0.75,0,0}
\definecolor{blue}{rgb}{0,0,0.75}
\definecolor{green}{rgb}{0,0.5,0}

\DeclareMathOperator{\tr}{tr}

\begin{document}

\title{Orientational correlations in active and passive nematic defects}

\author{D. J. G. Pearce}
\thanks{Authors have contributed equally}
\affiliation{Department of Mathematics, Massachusetts Institute of Technology, Cambridge, Massachusetts 02142, USA}
\affiliation{Departments of Biochemistry and Theoretical Physics, Universit\'e de Gen\'eve, 1205 Gen\'eve, Switzerland}
\author{J. Nambisan}
\thanks{Authors have contributed equally}
\affiliation{Department of Condensed Matter Physics, University of Barcelona, 08028 Barcelona, Spain}
\affiliation{School of Physics, Georgia Institute of Technology, Atlanta, Georgia 30332, USA}
\author{P. W. Ellis}
\affiliation{John A. Paulson School of Engineering and Applied Sciences, Harvard University, Cambridge, MA, 02138, USA}
\author{A. Fernandez-Nieves}
\email{a.fernandeznieves@ub.edu}
\affiliation{Department of Condensed Matter Physics, University of Barcelona, 08028 Barcelona, Spain}
\affiliation{School of Physics, Georgia Institute of Technology, Atlanta, Georgia 30332, USA}
\affiliation{ICREA-Institucio Catalana de Recerca i Estudis Avancats, 08010 Barcelona, Spain}
\author{L. Giomi}
\email{giomi@lorentz.leidenuniv.nl}
\affiliation{Instituut-Lorentz, Universiteit Leiden, P.O. Box 9506, 2300 RA Leiden, The Netherlands}

\begin{abstract}
We investigate the emergence of orientational order among $+1/2$ disclinations in active nematic liquid crystals. Using a combination of theoretical and experimental methods, we show that $+1/2$ disclinations have short-range antiferromagnetic alignment, as a consequence of the elastic torques originating from their polar structure. The presence of intermediate $-1/2$ disclinations, however, turns this interaction from anti-aligning to aligning at scales that are smaller than the typical distance between like-sign defects. No long-range orientational order is observed. Strikingly, these effects are insensitive to material properties and qualitatively similar to what is found for defects in passive nematic liquid crystals.
\end{abstract}

\maketitle

Topological defects are one of the hallmarks of liquid crystals and have represented a central research topic since Frank's pioneering work on nematic disclinations \cite{Frank:1958}. In nematics, these are point- or line-singularities where molecules have undefined orientation and around which the nematic director winds by an integer multiple of $\pi$ \cite{DeGennes:1993}. In planar nematic textures, it has long been known that, analogous to charged particles in two dimensions, disclinations interact with each other via long-ranged Coulomb-like forces arising from the distortion of the nematic director. As a consequence, like-sign disclinations repel, whereas oppositely-signed disclinations attract and eventually annihilate \cite{DeGennes:1993,Chaikin:1995}. 

By contrast, only recently has it become evident that most two-dimensional disclinations have a well-defined polarity, which affects how defects move and interact \cite{Vromans:2016}. Despite being a purely geometrical property of the director configuration, thus independent of the specific physical mechanisms governing the dynamics of the underlying nematic phase, such a polarity was first discussed in the context of active nematics \cite{Keber:2014,DeCamp:2015}, namely nematic liquid crystals consisting of self- or mutually-propelled rod-like molecules. In active nematics, polarity determines the propulsion direction of $+1/2$ disclinations \cite{Giomi:2013,Giomi:2014}, affects the attractive and repulsive interactions of defect pairs \cite{Kumar:2018,Pearce:2021}, and renders their motion periodic when confined on a sphere \cite{Keber:2014,Khoromskaia:2017}. Furthermore, polarity can be manipulated via inhomogeneous \cite{Ellis:2018,Pearce:2019a} or anisotropic \cite{Guillamat:2016,Guillamat:2017,Pearce:2019b,Pearce:2020} substrates. Perhaps even more remarkably, chaotic active nematics at the water-oil interface have been reported to exhibit long-ranged nematic order among the defects themselves, resulting from the alignment of the microscopic polarity of individual $+1/2$ disclinations over the length scale of the entire sample \cite{DeCamp:2015}. The physical origin of this behavior has, however, remained elusive, despite the efforts to decipher the mechanism behind this exotic example of {\em super} orientational order \cite{Putzig:2016,Doostmohammadi:2016,Oza:2016,Srivastava:2016,Kumar:2018,Shankar:2019,Patelli:2019,Thijssen:2020}.

In this Letter we investigate the mechanisms underpinning collective defect ordering in passive and active nematic liquid crystals. Using nematic hydrodynamics and experiments with microtubule-kinesin-based active nematics, we demonstrate that the elastic torques arising from the polarity of $+1/2$ disclinations drive the emergence of short-ranged antiferromagnetic alignment. Having an elastic origin, such an ordering effect occurs in passive and active nematics alike. However, in active nematics the continuous creation and annihilation of defects renders defect ordering stationary, whereas in passive nematics this occurs only as a transient phenomenon during defect coarsening. Furthermore, we demonstrate that $-1/2$ defects can mediate these orientational interactions by promoting ferromagnetic alignment at short distances.

\begin{figure}[t!]
\centering
\includegraphics[width=\columnwidth]{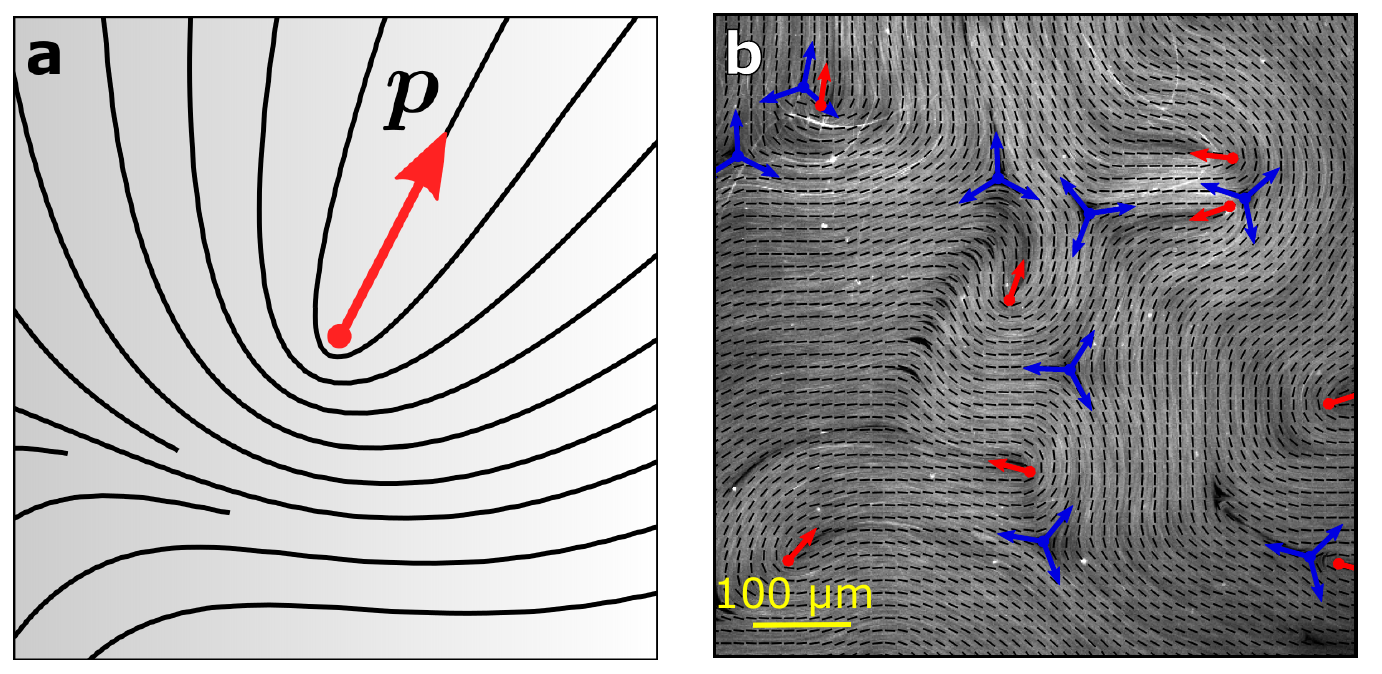}
\caption{\label{fig:illustration}(a) Schematic representation of a planar nematic featuring a $+1/2$ disclination with polarity $\bm{p}$. (b) Typical configuration of a microtubule-kinesin-based active nematic. $+ 1/2$ and $- 1/2$ disclinations are indicated with red and blue arrows.} 
\end{figure}

Let us consider a two-dimensional nematic liquid crystal whose average orientation is characterized by the director $\bm{n}=(\cos\theta,\sin\theta)$. In the presence of a disclination of strength $s=\pm 1/2$ located at the origin of the $xy-$plane and oriented in the direction $\bm{p}=(\cos\psi,\sin\psi)$, the local orientation is given by: $\theta = s\phi+(1-s)\psi$, with $\phi = \arctan(y/x)$ the polar angle \cite{Vromans:2016} (Fig. \ref{fig:illustration}a). Since the free energy of nematic liquid crystals is ${\rm O}(2)$ symmetric \cite{DeGennes:1993}, there is no preferential $\psi$ value for an isolated defect. Conversely, textures comprising multiple defects are sensitive to their relative orientation and attain the lowest energy configuration for specific alignment patterns \cite{Vromans:2016,Oza:2016,Tang:2017,Cortese:2018,Missaoui:2020,Pearce:2021}. In particular, pairs of $+1/2$ disclinations embedded in an otherwise defect-free nematic texture, tend to anti-align in order to minimize the system free energy~\cite{Vromans:2016,Oza:2016,Tang:2017,Cortese:2018,Missaoui:2020,Pearce:2021}. In passive and active nematic liquid crystals featuring multiple defects, one may expect these orientational interactions among pairs of $+1/2$ defects to cooperatively give rise to orientational order among the defects themselves, possibly leading to long-ranged defect ordering~\cite{DeCamp:2015}. 

\begin{figure}[h!]
\centering
\includegraphics[width=\columnwidth]{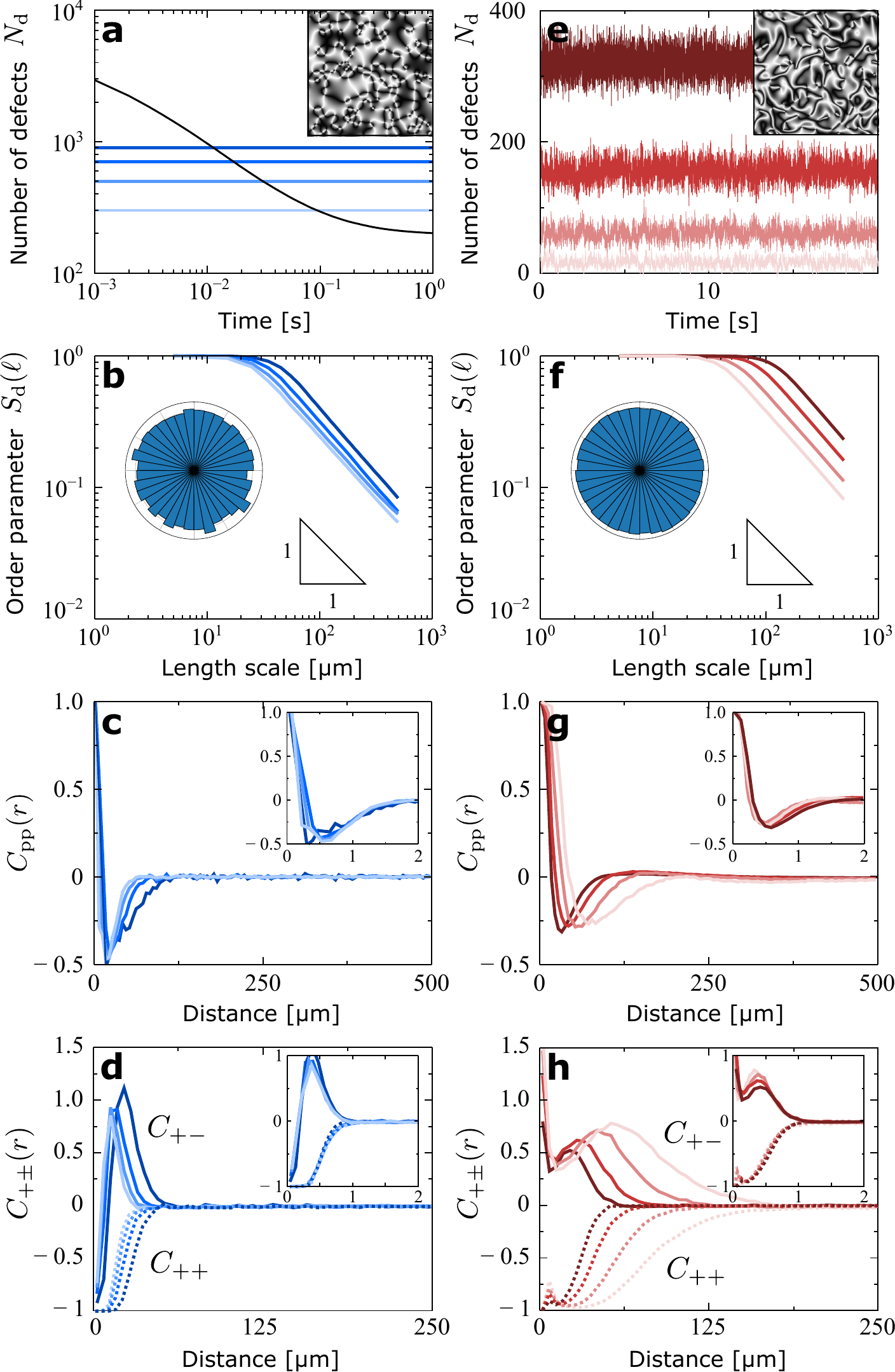}
\caption{\label{fig:hydrodynamics}Numerical Results. (a) Number of defects $N_{\rm d}$ versus time in coarsening passive nematics. The horizontal lines mark configurations featuring $N_{\rm d}=300,\,500,\,700,\,900$ defects. Inset: snapshot of a typical configuration. (b) Scale-dependent defect nematic order parameter, showing the characteristic scaling behavior of systems with no long- or quasi-long-ranged order: $S_{\rm d}(\ell)\sim \ell^{-1}$. Inset: polar histogram of the orientation of $N_{\rm d}=100$, $+1/2$ defects. (c) Polar correlation function $C_{\rm pp}(r)=\langle \bm{p}(\bm{r})\cdot\bm{p}(\bm{0}) \rangle$. Inset: the same function versus $r/r_{0}$, with $r_{0}\sim1/\sqrt{N_{\rm d}}$ the average defect spacing. The data collapse on the same master curve. (d) Topological charge density correlation functions for $C_{+-}$ ($C_{++}$), showing a local depletion (accumulation) of like-sign (opposite-sign) defects at short distances. Inset: the same functions versus $r/r_{0}$. (e-h) Same quantities as in (a-d) but for an active nematic whose dynamics are governed by Eqs. \eqref{eq:hydrodynamics}. The prominently different behaviors of $C_{+-}$ in the passive (d) and active (h) cases at $r \approx 0$ originate from the random creation of defect pairs and subsequent self-propulsion of +1/2 defects in active nematics, which temporarily results in $\pm1/2$ pairs at short distances. In all numerical simulations we have set $\lambda=0.1$, $\rho=10^{-2}\,{\rm kg}\,{\rm m}^{-2}$, $\eta=10^{-4}\,{\rm kg}\,{\rm s}^{-1}$ and $\alpha=\{-6.25,\,-12.50,\,-25.00,\,-50.00\}\,\times 10^{-2}\,{\rm kg}\,{\rm s}^{-2}$ \cite{SI}.
}	
\end{figure}

In order to test this hypothesis and decouple the effects of orientational elasticity and hydrodynamics, we start from the case of a two-dimensional passive nematic relaxing toward the minimum of the Landau-de Gennes free energy $F_{\rm LdG}~=~K/2\,\int {\rm d}A\,\left[|\nabla\bm{Q}|^{2}+1/(2\epsilon^{2})\tr\bm{Q}^{2}(\tr\bm{Q}^2-1)\right]$, with $\bm{Q}$ the two-dimensional nematic tensor \cite{DeGennes:1993} and $\epsilon$ a length scale setting the defect core radius. Thus
\begin{equation}\label{eq:relaxation}
\frac{\partial \bm{Q}}{\partial t} = \frac{1}{\gamma}\,\bm{H}\;,
\end{equation}
where $\gamma$ is the rotational viscosity and  $\bm{H}=-\delta F/\delta \bm{Q}$. Eq.~\eqref{eq:relaxation} is numerically integrated using finite differences on a periodic square domain of size $L$, subdivided in $256 \times 256$ collocation points, starting from a random configuration (Fig. \ref{fig:hydrodynamics}a inset). In all our simulations we set $K=10^{-8}\,{\rm N}\,{\rm m}$, $\gamma=10^{-3}\,{\rm kg}\,{\rm s}^{-1}$ and $\epsilon=10^{-2}\,L$ (see Ref. \cite{SI} for details about the choice of material parameters).

As the system is allowed to relax, pairs of $\pm 1/2$ defects attract and annihilate toward a defect-free and uniformly aligned configuration and the number of defects $N_{\rm d}$ then decreases in time (Fig. \ref{fig:hydrodynamics}a). To characterize defect ordering, we store configurations at four different times, corresponding to $N_{\rm d}=300,\,500,\,700,\,900$ defects (horizontal lines in Fig. \ref{fig:hydrodynamics}a). We then measure the probability distribution of the defect local orientation $\psi$ to find that it is prominently uniform in the interval $0\le \psi \le 2\pi$ (inset in Fig. \ref{fig:hydrodynamics}b), indicating the absence of long-range polar or nematic order. 

To verify whether the elastic interactions give rise to quasi-long-range nematic order, we measure the scale-dependent nematic order parameter $S_{\rm d}(\ell)=[\langle \cos 2\psi \rangle_{\ell}^{2}+\langle \sin 2\psi  \rangle_{\ell}^{2}]^{1/2}$, where $\langle \cdots \rangle_{\ell}$ denotes a spatial average over a square sub-domain of size $\ell$. The procedure is repeated 100 times to obtain the statistically-averaged $S_{\rm d}(\ell)$ values displayed in Fig. \ref{fig:hydrodynamics}b.
In case of long-ranged order, this parameter converges to a finite limit for large $\ell$. Conversely, in quasi-long-ranged ordered samples, $S_{\rm d}(\ell) \sim \ell^{-\eta_{\rm d}/2}$, with $\eta_{\rm d}<1/4$ a positive non-universal exponent~\cite{Udink:1987}. Finally, for randomly oriented defects $S_{\rm d}(\ell)\sim \ell^{-1}$. The data presented in Fig. \ref{fig:hydrodynamics}b is consistent with this latter scenario, thus indicating that when passive defects coarsen from a random configuration of the nematic tensor they exhibit neither long-ranged nor quasi-long-ranged orientational order. In spite of this, the orientational interactions among the defects leave a well defined signature in the orientational correlation function $C_{\rm pp}(r)=\langle \bm{p}(\bm{r})\cdot\bm{p}(\bm{0})\rangle$ displayed in Fig. \ref{fig:hydrodynamics}c. Before vanishing at large distances, this correlation function exhibits a prominent minimum at small $r$ values, indicating the preference for local antiferromagnetic alignment. Remarkably, all curves collapse to the same master curve upon rescaling distances by the average defect spacing $r_{0}\sim 1/\sqrt{N_{\rm d}}$ (Fig. \ref{fig:hydrodynamics}c inset).  

Similarly, Fig. \ref{fig:hydrodynamics}d shows the topological charge density correlation function 
\begin{equation}
C_{ab}(r)=\frac{\langle \rho_{a}(\bm{r})\rho_{b}(\bm{0}) \rangle}{\langle \rho_{a}(\bm{0}) \rangle \langle \rho_{b}(\bm{0}) \rangle} - 1\;,
\end{equation}
where $(a,b)\in \{+,-\}$ and $\rho_{\pm}$ represent the topological charge density of $\pm 1/2$ disclinations. For small $r$ values, $C_{++}(r) \approx -1$ (blue dotted lines), indicating that the space surrounding a positive defect is depleted of like-sign defects. Starting from $r/r_0 \approx 0.25$ (Fig. \ref{fig:hydrodynamics}d inset), the same function exhibits a monotonic increase, until plateauing at  $C_{++}(r) \approx 0$ for $r/r_0 > 1$. The same behavior is found for $C_{--}$~\cite{SI}. By contrast, $C_{+-}(r)$ (blue lines) exhibits a prominent peak at $r/r_0 \approx 0.25$, indicating a local concentration of oppositely charged defects. This is again followed by a rapid convergence toward $C_{+-}(r) \approx 0$ for $r/r_0 > 1$ (Fig. \ref{fig:hydrodynamics}d inset). The combination of these results demonstrates that, analogous to Debye screening in electrolytes and consistently with the classic Coulomb gas picture of disclinations in liquid crystals \cite{Nelson:2002}, positive defects are surrounded by clouds of negative defects and vice versa, but that they are not endowed of positional order of any kind.

Next, we explore the effect of activity on defect ordering. As demonstrated in Refs. \cite{Giomi:2013,Giomi:2014} and later recovered in various experiments on active nematics of biological origin \cite{Keber:2014,Duclos:2017,Kawaguchi:2017,Saw:2017,Blanch-Mercader:2018,Lemma:2019}, in the presence of active stresses, the strong distortion introduced by topological defects gives rise to hydrodynamic flows, whose structure depends solely on the local geometry of the nematic director in the proximity of the core. These active flows, in turn, can affect the relative alignment of the defects via hydrodynamic torques~\cite{Pearce:2019a}. In order to test whether these activity-driven hydrodynamic torques influence defect ordering, we have numerically integrated the hydrodynamic equations of an incompressible two-dimensional active nematic (e.g. Ref.~\cite{Giomi:2012}):
\begin{subequations}\label{eq:hydrodynamics}
\begin{gather}
\frac{D \bm{Q}}{D t} = \lambda S \bm{u}+\bm{Q}\cdot\bm{\omega}-\bm{\omega}\cdot\bm{Q}+\frac{1}{\gamma}\,\bm{H}\;, \\
\rho\,\frac{D \bm{v}}{D t} = \eta \nabla^{2} \bm{v} + \nabla\cdot(\bm{\sigma}^{\rm p}+\alpha\bm{Q})\;, \quad \nabla\cdot\bm{v}=0\;. 
\end{gather}	
\end{subequations}
Here $D/Dt = \partial/\partial t+\bm{v}\cdot\nabla$ is the material derivative, $\lambda$ is the flow-aligning parameter of the nematic fluid, $\bm{u}=[\nabla\bm{v}+(\nabla\bm{v})^{\rm T}]/2$ and $\bm{\omega}=[\nabla\bm{v}-(\nabla\bm{v})^{\rm T}]/2$ are, respectively, the strain-rate and vorticity tensors, $\rho$ the density, here assumed to be constant, $\eta$ the shear viscosity and $\bm{\sigma}^{\rm p}=-P\mathbb{1}-\lambda S \bm{H}+\bm{Q}\cdot\bm{H}-\bm{H}\cdot\bm{Q}$, with $P$ the pressure, is the passive reactive stress tensor. The final term in Eq. (\ref{eq:hydrodynamics}b) is the active stress originating from local contractile ($\alpha>0$) or extensile ($\alpha<0$) forces exerted by the active nematogens. 

When the system size is much larger than the intrinsic length scale $\ell_{\rm a}=\sqrt{K/|\alpha|}$, resulting from the balance of active and passive torques, two-dimensional active nematics self-organize in a turbulent-like steady-state comprising a stationary density ($\sim 1/\ell_{\rm a}^{2}$) of $\pm 1/2$ defects \cite{Giomi:2015} (Fig. \ref{fig:hydrodynamics}e inset). Analogous to the case of passive defects coarsening from a highly defective configuration, the scale dependent nematic order parameter of $+1/2$ defects decays as $S_{\rm d}(\ell)\sim\ell^{-1}$, indicating the lack of long- or quasi-long-ranged order (Fig. \ref{fig:hydrodynamics}f). Instead, defect ordering emerges again in the form of short-ranged antiferromagnetic alignment (Fig. \ref{fig:hydrodynamics}g), coupled to local Debye-like screening of the topological charge (Fig. \ref{fig:hydrodynamics}h). 

A comparison between our results for the passive (Fig. \ref{fig:hydrodynamics}a-d) and active (Fig. \ref{fig:hydrodynamics}e-h) cases suggests that, unlike what was previously thought, defect ordering ultimately originates from passive mechanisms and can be found in passive and active liquid crystals alike. This consideration is further supported by the fact that, as demonstrated by Fig. \ref{fig:hydrodynamics}c-d and \ref{fig:hydrodynamics}g-h, defect spatial and orientational correlations support a common scaling variable with respect to data collapse. The latter scenario is common in near-equilibrium systems subject to a gradient-descent dynamics, such as that embodied in Eq. \eqref{eq:relaxation}, but generally violated in far from equilibrium systems \cite{Cross:1995}. Our data therefore strengthens the idea that, despite chaotic active nematics representing one of the most iconic examples of out of equilibrium systems, many of their structural features, including the statistics of vortices \cite{Giomi:2015} and topological defects, follow from the same passive mechanisms found in equilibrium and near-equilibrium systems.

Finally, we note that the magnitude of the orientational correlation function at the antiferromagnetic minimum is smaller for active nematics than for passive nematics (Figs. \ref{fig:hydrodynamics}c and g). Hence, active flows collectively hinder defect ordering rather than enhance it. This can be intuitively understood by noticing that, whereas the relaxational dynamics of passive nematics is solely dictated by elastic interactions, in active nematics director orientations are randomized by the flow due to the persistent injection of active stress. This effect contrasts the ordering effect of the elastic interactions, but without completely destroying it, as increasing activity leads to an increase in defect density, which, in turn, enhances elastic interactions by decreasing the distance between defects. In addition, our numerical data shows no evidence of a correlation between extensile (contractile) activity and aligning (anti-aligning) interactions between defects \cite{SI}, as that reported in Ref. \cite{Shankar:2018}.

\begin{figure}[t!]
\centering
\includegraphics[width=\columnwidth]{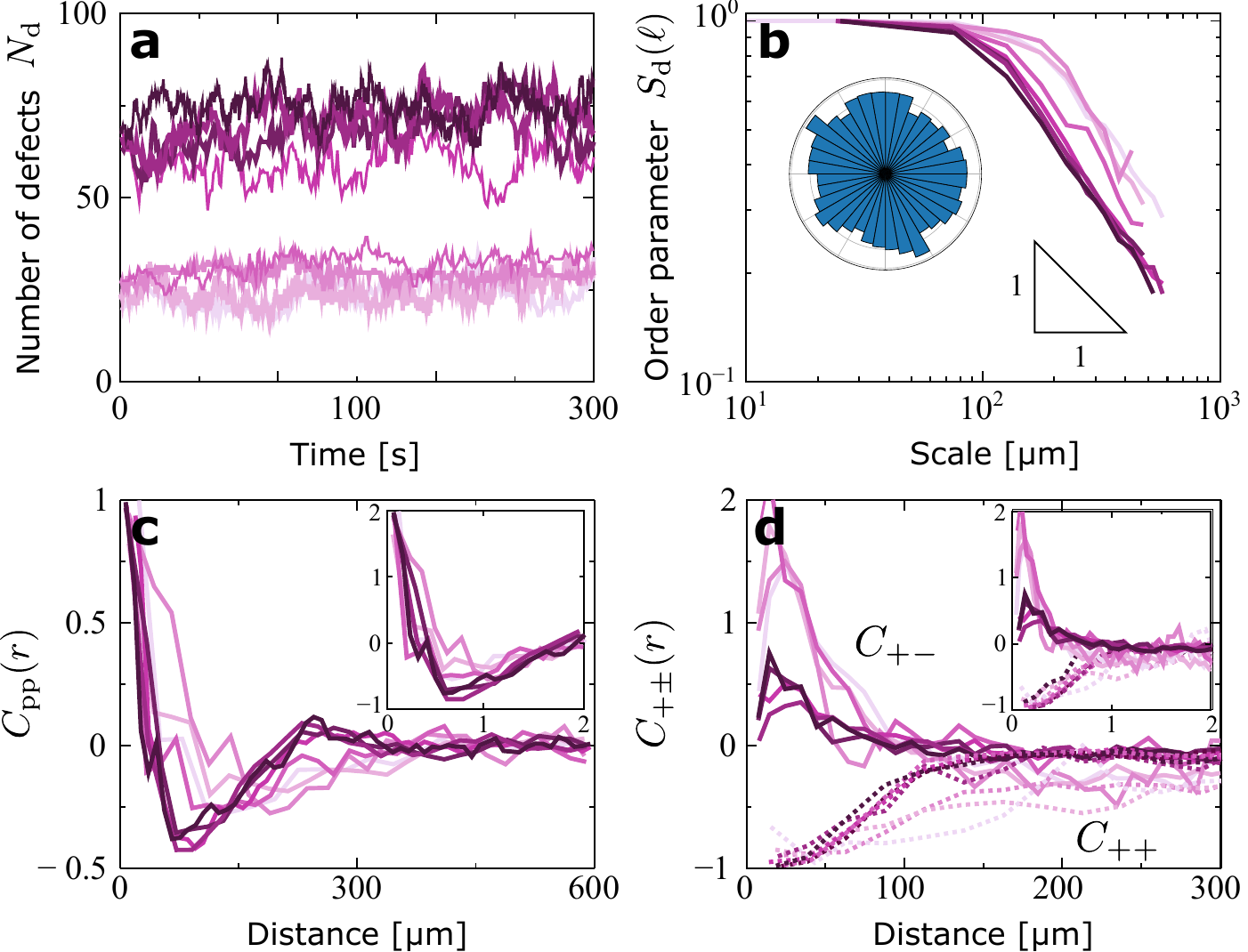}
\caption{\label{fig:exptfig}Experimental Results. (a) Number of defects versus time in a microtuble/kinesin-based active nematic. Each color represents a different experimental sample, grouped into two sets with different $N_{\rm d}$. (b) Scale-dependent nematic order parameter of the $+1/2$ defects. Inset: polar histogram of the orientation of the $+1/2$ defects. (c) Polar correlation function $C_{\rm pp}=\langle \bm{p}(r)\cdot\bm{p}(0)\rangle$. (d) Correlation functions $C_{+-}$ and $C_{++}$. Insets: same functions versus $r/r_0$.}	
\end{figure}

To asses the significance of our predictions, we carry out experiments on active nematic suspensions of microtubules (MTs) \cite{Sanchez:2012}. The system is driven out of equilibrium by the action of kinesin-streptavidin motor protein complexes, which induce relative motion between the MT bundles utilizing adenosine triphosphate (ATP) as the energy source. A typical snapshot of a confocal frame is shown in Fig. \ref{fig:illustration}b. We perform several experimental replicas for two different activities, which we vary through the ATP concentration ($72\,\mu$M and $144\,\mu$M). We then extract defect positions and orientations using computer vision techniques \cite{Ellis:2020,SI}. Consistent with our numerical simulations, we find that the average number of defects for each experiment is constant over time (Fig. \ref{fig:exptfig}a), while defect polarity is isotropically distributed (Fig. \ref{fig:exptfig}b inset). Moreover, both the scale-dependent nematic order parameter (Fig. \ref{fig:exptfig}b) and the correlation functions (Fig. \ref{fig:exptfig}c,d) confirm that the orientational order among $+1/2$ defects is short-ranged and that the order parameter approximately decays as $S_{\rm d}(\ell)\sim\ell^{-1}$. Scaling by the average defect spacing $r_0$, the curves approximately collapse onto the same master curve (inset of Figs. \ref{fig:exptfig}c,d).

\begin{figure}[t]
\centering
\includegraphics[width=\columnwidth]{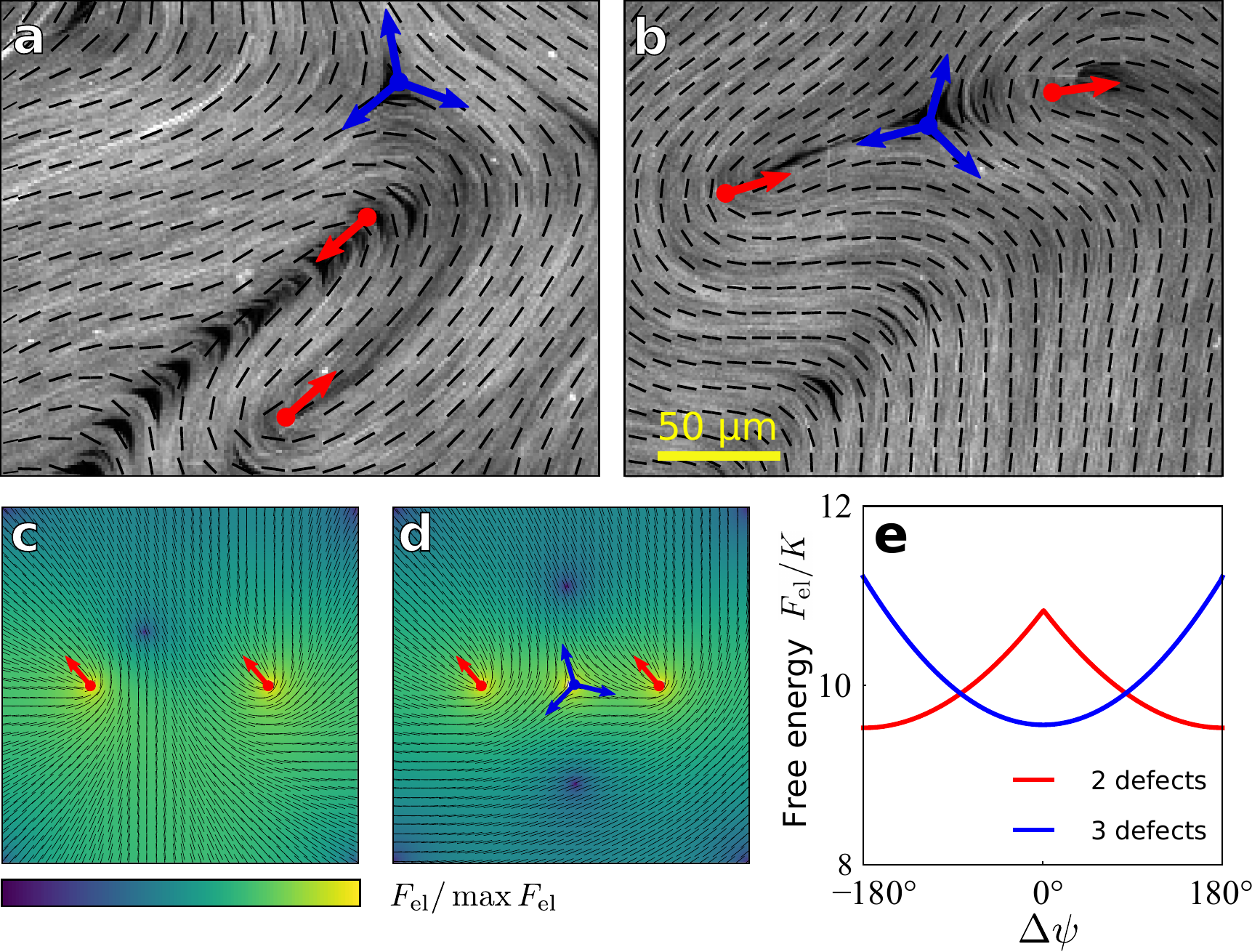}
\caption{\label{fig:defectconfigs}(a), (b) Snapshots illustrating the (a) antiferromagnetic alignment between $+1/2$ defects, and (b) ferromagnetic alignment between $+1/2$ defects in the in the presence of an intermediate $-1/2$ defect. These effects determine the orientational correlation, for $r<r_{0}$, and anti-correlation, for $r>r_{0}$, in Figs. \ref{fig:hydrodynamics}g and \ref{fig:exptfig}c. (c,d) Example of configurations consisting of two $+1/2$ disclinations in the absence (c) and in the presence (d) of an intermediate $-1/2$ disclination. (e) Elastic free energy of the configurations displayed in panels (c) and (d) as a function of the angular separation $\Delta\psi$ between $+1/2$ defects. These are energetically favored to be anti-aligned while separated by a defect free patch (red curve) and aligned in the presence of an intermediate $-1/2$ defect (blue curve).}
\end{figure}

Importantly, we observe that at long distances ($r > r_{0}$), $+1/2$ defects are randomly aligned and have nearly vanishing correlation, as expected in the absence of orientational order. For $r \approx r_{0}$, however, the defects anti-align on average (Fig. \ref{fig:defectconfigs}a). Yet, as it was also noticed in actomyosin films \cite{Kumar:2018}, the presence of oppositely charged $-1/2$ defects can mediate this short-range interaction, eventually favoring the polar alignment between two $+1/2$ defects at distances shorter than $r_{0}$ (Fig. \ref{fig:defectconfigs}b). To further shed light on this crossover, we have numerically computed the elastic free energy $F_{\rm el}=K/2\int {\rm d}A\,|\nabla\bm{Q}|^{2}$ of two configurations consisting of two $+1/2$ disclinations in the absence (Fig. \ref{fig:defectconfigs}c) and in the presence (Fig. \ref{fig:defectconfigs}d) of an intermediate $-1/2$ disclination. The free energy is plotted in Fig. \ref{fig:defectconfigs}e as a function of the angle $\Delta\psi$ between $+1/2$ defects. Consistent with our experimental observations, the $+1/2$ defects are energetically favored to be anti-aligned while separated by a defect free patch and aligned in the presence of a $-1/2$ defect. Thus, we conclude that $-1/2$ defects mediate the orientational interactions between $+1/2$ defects at distances $r<r_{0}$, by promoting alignment. 

In conclusion, we have demonstrated that in both passive and active nematics $+ 1/2$ disclinations exhibit short-ranged orientational correlations in the form of antiferromagnetic alignment, at distances comparable to the mean inter-defect spacing, and ferromagnetic alignment at even smaller distances. The latter is enabled by the presence of nearby oppositely charged excitations, which screen the repulsion between like-charge defects, in a way reminiscent of ionic screening in electrolyte solutions. Crucially, we find no signature of long or quasi-long-ranged order among defects. Our finding is consistent with other experimental studies using microtubule-kinesin active nematics~\cite{SI}, including Ref.~\cite{DeCamp:2015}, where the order parameter $S_{\rm d}(\ell)$ has likely been overestimated. This lack of long-ranged order could nonetheless play a functional role in biological active nematics, such as specific embryonic tissues, where defects have been likened to ``topological morphogens'' \cite{MaroudasSacks:2021,Hoffmann:2021}. In this respect, the absence of a preferential direction, resulting from an organism-wide breakdown of rotational symmetry, could guarantee this mechanism a certain amount of versatility. Finally, our results clearly show that the observed effects have an elastic origin even in the active case, where the nematogens are driven out-of-equilibrium by local energy-input. 

\acknowledgements

We are thankful to Zvonimir Dogic for useful discussions and to Berta Martinez-Prat, Jordi Ign\'es-Mullol and Francesc Sagu\'es for help with active-nematic preparation and sharing their facilities. This work was supported by the Netherlands Organization for Scientific Research (NWO/OCW), as part of the Frontiers of Nanoscience program and the Vidi scheme, and by MCIU/AEI/FEDER,UE (PGC2018-097842-B-I00).

\end{document}